\documentclass[amsmath,amssymb,pra,twocolumn]{revtex4}
\usepackage[cp1251]{inputenc}
\usepackage[english]{babel}
\usepackage{color}
\usepackage{graphicx}
\usepackage{dcolumn}
\usepackage{braket}
\usepackage{amsmath}

\usepackage{natbib}

\begin{document}

\title[Hong-Ou-Mandel effect in terms of the temporal wave function]
{Hong-Ou-Mandel effect in terms of the temporal biphoton wave function with two arrival-time variables.}

\author{M.V. Fedorov$^{1, 3\,*}$, A.A. Sysoeva$^{1, 2}$, S.V. Vintskevich$^{1, 2}$, D.A. Grigoriev$^{1, 2}$}
\address{$^{1}$ A.M. Prokhorov General Physics Institute, Russian Academy of Sciences,38 Vavilov st., Moscow, 119991, Russia.  \\
$^{2}$ Moscow Institute of Physics and Technology, Dolgoprudny, Moscow Region, Russia\\
$^{3}$ National Research University ``Higher School of Economics", 20 Myasnitskaya st,
Moscow, 101000, Russia\\
$^*{\rm fedorovmv@gmail.com}$}

\date{\today}

\begin{abstract}
The well-known Hong-Ou-Mandel effect is revisited. Two physical reasons are discussed for the effect to be less pronounced or even to disappear: differing polarizations of photons coming to the beamsplitter and delay time of photons in one of two channels. For the latter we use the concepts of biphoton frequency and temporal wave functions depending, correspondingly, on two frequency continuous variables of photons and on two time variables $t_1$ and $t_2$ interpreted as the arrival times of photons to the beamsplitter. Explicit expressions are found for the probability densities and total probabilities for photon pairs to be split between two channels after the beamsplitter and to be unsplit, when two photons appear together in one of two channels.

\end{abstract}


\maketitle


\section{Introduction}

\begin{figure}[h]
\centering
\includegraphics[width=5 cm]{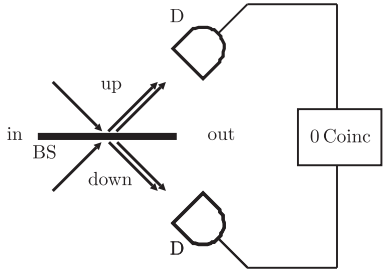}
\caption{{\protect\footnotesize {A scheme for observing the HOM effect. $\rm BS$ is the beamsplitter. $D$ refers to detectors}.}}\label{Fig1}
\end{figure}

The Hong-Ou-Mandel (HOM) effect \cite{Hong} is well known, it finds many practical
applications and is discussed by many authors in many publications \cite{Rubin,Zeilinger, Giuzeppe, Pan, Abouraddy, Ulanov}. In the HOM effect two photons are assumed to be sent to the $50-50\%$ beam-splitter (BS)
under the angles $45^\circ$ to the horizontal BS-plane, one of photons coming from the upper,
and another one from the lower halfplanes (Fig.\ref{Fig1}). Below photons and creation
operators corresponding to propagation in the upper and lower half-planes will be indexed
correspondingly by the letters $``u"$ and $``d"$  (the latter from $down$ which is not
related to the down-propagation directions but rather to location of photons in the lower halfplane which is down compared to the upper halfplane).
If both photons have the same given frequency and coinciding polarizations, the incoming state vector is given by
\begin{equation}
\label{in-state}
    \ket{\Psi_{in}}=a_u^\dag a_d^\dag\ket{0}=\ket{1_u,1_d},
\end{equation}
where $a_u^\dag$ and $a_d^\dag$ are the photon creation operators for photons in the up- and down- halfplanes before reflection from or propagation through the BS.

Let the BS transformation matrix have the simplest form
$\displaystyle\frac{1}{\sqrt{2}}\left(\begin{matrix}\;\;1\,\, 1\\-1\,\,1\end{matrix}\right)$. Then the transformation rules for the creation operators are
\begin{equation}
 \nonumber
    a^{\dag}_u\rightarrow\displaystyle\frac{1}{\sqrt{2}}\left(a_u^\dag-a_d^\dag\right)\,{\rm and}\;\,
    \label{a-dag-transformed}
    a_d^\dag\rightarrow\displaystyle\frac{1}{\sqrt{2}}\left(a_u^\dag+a_d^\dag\right),
\end{equation}
which gives the following expression for the biphoton state vector after propagation/reflection through/from the BS
\begin{equation}
 \nonumber
   {\ket{\Psi_{out}}=\frac{1}{2}
  \left(\,a_u^\dag-a_d^\dag\right) \left(\,a_u^\dag+a_d^\dag\right)\ket{0}
  =}\\
\end{equation}
\begin{equation}
 \frac{1}{2}\displaystyle(\,a_u^{\dag^{\,2}}-a_d^{\dag^{\,2}})\ket{0}=
 \frac{1}{\sqrt{2}}(\ket{2_u}-\ket{2_d}).
 \label{HOM}
\end{equation}
These expressions correspond to the ideal HOM effect. Transformed by BS, photon pairs propagate  unsplit either in up- or in down halfplanes, and the probability of observing pairs split between the up- and down-channels equals zero. In experiments this can be seen as a zero signal of coincidence measurements between these two channels, as shown in Fig. \ref{Fig1}.

The given derivation in terms of the photon's creation operators is very simple and compact. But it does not demonstrate explicitly the interference origin of the HOM effect, which can be demonstrated more clearly in the wave-function formalism. The one-photon wave functions corresponding to operators $a_u^\dag$ and $a_d^\dag$ can be written as
$$\psi_u=\bra{\xi}a_u^\dag\ket{0}=\delta_{\xi,u}=\left(\begin{matrix}1\\ 0\end{matrix}\right)\;{\rm and}$$
$$\psi_d=\bra{\xi}a_d^\dag\ket{0}=\delta_{\xi,d}=\left(\begin{matrix}0\\1\end{matrix}\right),$$
where $\xi$ is a variable differing $up$ and $down$ halfplanes, $\delta$ denotes the Kronecker symbol, and in the columns the upper and lower lines correspond, respectively, to the $up$- and $down$-halfplanes. In the column-representation the biphoton wave function, corresponding to the state vector of Eq. (\ref{in-state}) is given by
\begin{equation}
 \nonumber
\Psi_{in} \equiv \Psi_+=
\end{equation}
\begin{equation}
\frac{1}{\sqrt{2}}\left[\left(\begin{matrix}1\\ 0\end{matrix}\right)_1
\left(\begin{matrix}0\\1\end{matrix}\right)_2+\left(\begin{matrix}0\\ 1\end{matrix}\right)_1\left(\begin{matrix}1\\ 0\end{matrix}\right)_2\right]
 \label{directional wave function}
\end{equation}
where the indices $1$ and $2$ correspond to numbers of not shown explicitly variables $\xi_1$ and $\xi_2$. The biphoton wave function (\ref{directional wave function}) is symmetrized with respect to the transposition of variables $1\leftrightarrows 2$, which is obligatory for photons as indistinguishable particles obeying the Bose-Einstein statistics; $\Psi_+$ in Eq. (\ref{directional wave function}) is the notation of one of the Bell states for biphotons with variables $\xi_1$ and $\xi_2$. After the BS transformation the first and second terms in the sum of Eq. (\ref{directional wave function}) yield
\begin{equation}
 \displaystyle\frac{1}{2}\Big(\Phi_- +\Psi_-\Big)\; {\rm and}\;
  \displaystyle\frac{1}{2}\Big(\Phi_- -\Psi_-\Big),
\label{transformed w.f.}
\end{equation}
where $\Phi_-$ and $\Psi_-$ are notations of two other Bell states
\begin{equation}
 \label{Bell Phi-}
 \Phi_-=\frac{1}{\sqrt{2}}\left[\left(\begin{matrix}1\\ 0\end{matrix}\right)_1
\left(\begin{matrix}1\\ 0\end{matrix}\right)_2-\left(\begin{matrix}0\\ 1\end{matrix}\right)_1
\left(\begin{matrix}0\\ 1\end{matrix}\right)_2\right]
\end{equation}
and
\begin{equation}
 \label{Bell Psi-}
 \Psi_-=\frac{1}{\sqrt{2}}\left[\left(\begin{matrix}1\\ 0\end{matrix}\right)_1
\left(\begin{matrix}0\\ 1\end{matrix}\right)_2-\left(\begin{matrix}0\\ 1\end{matrix}\right)_1\left(\begin{matrix}1\\ 0\end{matrix}\right)_2\right] .
\end{equation}
The function $\Phi_-$ is symmetric with respect to the transposition $1\rightleftarrows 2$ and it characterizes unsplit photon pairs, whereas the Bell state $\Psi_-$ is antisymmetric and it would characterize pairs split between the $up-$ and $down-$channels if it would exist in the total wave function. But asymmetry of $\Psi_-$ indicates clearly that such state cannot be present in the total wave function and, indeed, it disappears in the sum of two terms in Eq. (\ref{transformed w.f.}), which gives finally
\begin{equation}
 \label{final w.f.}
 \Psi_{out}=\Phi_-
\end{equation}
in agreement with the result of Eq. (\ref{HOM}).

As it's clear from the given derivation, cancelation of the term $\Psi_-$ in the total wave function $\Psi_{out}$, as well as the existence of the HOM effect itself, occurs owing to interference of contributions from two terms in the incoming wave function $\Psi_{in}$ (\ref{directional wave function}), which illustrates also importance of symmetrization of biphoton wave functions.

\section{Biphoton states with differing polarizations of photons}

\begin{figure}[h]
\centering
\includegraphics[width=8 cm]{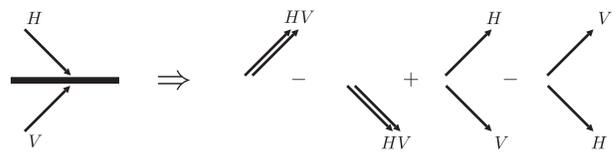}
\caption{{\protect\footnotesize {Equally probable four outgoing distributions of photon after BS for the incoming state $a_{u,\,H}^\dag a_{d,\,V}^\dag\ket{0}$ }.}}\label{Fig2}
\end{figure}

There are many reasons for which the ideal HOM effect can become less perfect or eliminated at all. One of such reasons is a possible difference of polarizations of incoming photons. In a general case, let $\alpha$ and $\beta$ be angles between linear polarizations (with respect to the horizontal one), correspondingly, of $up$- and $down$- incoming photons. Then the initial biphoton state vector takes the form
\begin{equation}
    \nonumber
    \ket{\Psi_{in}^{\alpha,\beta}}=a_{u,\,\alpha}^\dag a_{d,\,\beta}^\dag\ket{0}= (\cos\alpha\, a_{u,\,H}^\dag+\sin\alpha\, a_{u,\,V}^\dag)
\end{equation}
\begin{equation}
\times(\cos\beta\, a_{d,\,H}^\dag+\sin\beta\, a_{d,\,V}^\dag)\ket{0},
   \label{in-state-alpha-beta}
\end{equation}
where the indices $H$ and $V$ correspond to the horizontal and vertical polarizations. A simple algebra gives the following expression for a part of the state vector arising after interaction with the BS and corresponding to biphoton pairs split between the  $up$- and $down$-channels
\begin{equation}
\nonumber
\ket{\Psi_{out,\, {\rm split}}^{\alpha,\,\beta}}=
\end{equation}
\begin{equation}
  \label{split,alpha-beta}
-\frac{\sin(\alpha-\beta)}{2} \Big(\ket{1_{u,\,H};1_{d,\,V}}-\ket{1_{u,\,V};1_{d,\,H}}\Big).
\end{equation}
The corresponding total probabilities for biphoton pairs to be split and unsplit are given by
\begin{equation}
 \label{split}
  w_{\rm split}=\frac{1}{2}\sin^2(\alpha-\beta)\quad\quad
\end{equation}
\begin{equation}
  \label{unslplit}
  w_{\rm unsplit}=1-\frac{1}{2}\sin^2(\alpha-\beta).
\end{equation}
In particular, at $|\alpha-\beta|=\frac{\pi}{2}$ Eqs. (\ref{split}) and (\ref{unslplit}) give      $w_{\rm split}=w_{\rm unsplit}=\frac{1}{2}$. This means that in the case of orthogonal polarizations of incoming $up$- and $down$-photons the HOM effect completely disappears, and two incoming photons behave as absolutely independent particles, each of which is reflected or transmits the BS accidentally and with equal probabilities $1/2$. Such situation occurs, for example, in the case $\alpha=0$ and $\beta=\pi/2$, i.e., for the state $\ket{\Psi_{in}}=a_{u,\,H}^\dag a_{d,\,V}^\dag\ket{0}$. For this state the BS-transformation yields the state $\Psi_{out}$ given by
\begin{equation}
 \label{uH-dV}
\frac{1}{2}\Big(a_{uH}^\dag a_{uV}^\dag-a_{dH}^\dag a_{dV}^\dag+a_{uH}^\dag a_{dV}^\dag-a_{uV}^\dag a_{dH}^\dag\Big)\ket{0},
\end{equation}
which corresponds to four equally probable distributions of photons shown in Fig. \ref{Fig2}

On the other hand, it's interesting and important to notice that the split-pair part of the state vector of outgoing photons (\ref{split,alpha-beta}) is an odd function of the difference $\alpha-\beta$, and for the incoming state with reversed polarizations it changes sign:
\begin{equation}
 \label{reversed}
 \ket{\Psi_{out,\, {\rm split}}^{\beta,\,\alpha}}=-\ket{\Psi_{out,\, {\rm split}}^{\alpha,\,\beta}}.
\end{equation}
Hence, for the coherent superpositions (sum) of the state (\ref{in-state-alpha-beta}) and the same state with reversed polarizations their contributions to the split-pair part of the outgoing photons cancel each other, and their sum turns zero. This means that in the case of a state
\begin{equation}
 \label{coherent sum}
  \ket{{\widetilde\Psi}_{in}}=  \frac{1}{\sqrt{2}}\left(\ket{\Psi_{in}^{\alpha,\,\beta}}+\ket{\Psi_{in}^{\beta,\,\alpha}}\right)
\end{equation}
the ideal HOM effect restores owing to the interference cancelation of terms destroying this effect from two terms on the right-hand side of Eq. (\ref{coherent sum}). In particular, such an ideal HOM effect occurs for the state with uncertain horizontal or vertical polarization in the $up-$ and $down$-channels
\begin{equation}
  \ket{{\widetilde\Psi}_{in}}_{HV}=
  \label{H-V}
  \frac{\ket{1_{u,H},1_{d,V}}+\ket{1_{u,V},1_{d,H}}}{\sqrt{2}}.
\end{equation}
Cancelation of `wrong' contributions to the outgoing states from two terms in the sums of Eqs. (\ref{coherent sum}), (\ref{H-V}) illustrates once again an interference origin of the HOM effect.

\section{Time delay of photon arrivals}
Another reason for destroying the HOM effect can be related with the time delay of incoming photons in one of the channels, e.g., in the upper one. The simplest way for providing such time delay is lengthening the crystal-BS pathway for photons in the upper channel compared to the lower one. If $\Delta l$ is the lengthening distance, then the temporal delay of photons coming to BS is $\Delta t=\Delta l/c$. For photons with a given frequency $\omega$ such time delay results in multiplication of the upper-channel creation operator $a_u^\dag$ by the  phase factor $e^{i\omega\Delta t}$. But, evidently, this change will not affect the HOM effect at all because the phase factor will go through all transformations (\ref{in-state})-(\ref{HOM}) and will not add anything to the zero coincidence signal between the $up$ and $down$ channels in the transformed biphoton beam. Thus, in such simple model the ideal HOM effect exists at any values of the time delay  $\Delta t$, though in reality this cannot be true. The reason is in the assumption of a given value of the photon frequency. In this approximation photon wave functions are infinitely long in space and time, and lengthening of a pathway for some photons does not affect interference condition, because for any values of $\Delta t$ photons in one channel always find counterpart photons in the other channel to interfere with. Hence, for seeing the delay-dependent limitations of the ideal HOM effect, one has to take into account photon distributions in frequencies and related to this finiteness of the photon wave-packet lengths and durations. Here such consideration is based on the results of the work \cite{Mikhailova} slightly generalized for the noncollinear case.

Specifically, we consider the regime of the noncollinear type-I spontaneous parametric down conversion (SPDC) degenerate with respect to central frequencies of photons. Moreover, we do not consider angular distributions of photons by assuming only that in some given plane $(xz)$ containing the pump-propagation $z$-axis one of photons in each pair propagates at some angle $\theta_0$ to the $z$-axis, and the other one at the angle $-\theta_0$ ($up$ and $down$). As for the frequency biphoton wave function, we take it in the form found in the work \cite{Mikhailova} and consider only the case of long pump pulses (not less than 10 ps in accordance with estimates of Ref. \cite{Mikhailova}), when the frequency-dependent part of the wave function is given by
\begin{equation}
 \Psi^{({\rm freq})}\propto e^{i\frac{\tau_p^2(\omega_1+\omega_2-\omega_0)^2}{2}}
{\rm sinc}\left[\frac{LB}{2c\omega_0}(\omega_1-\omega_2)^2\right],
\end{equation}
where $\omega_1$ and $\omega_2$ are frequencies of emitted photons, $\omega_0$ is the central frequency of the pump, $\tau_p$ is the pump-pulse duration, ${\rm sinc}(x)=\sin x/x$, $L$ is the crystal length, $B$ is the dispersion constant $B=c(\omega_0/4)k_1^{\prime\prime}$, and $k_1^{\prime\prime}$ is the second-order derivative (over frequency) of the emitted-photon wave vector.

For simplicity, the sinc-function is modeled by the Gaussian function, ${\rm sinc}(x^2)\rightarrow \exp(-0.357x^2)$, directional parts of the symmetrized total wave function are characterized by the two-line columns, and the delay- dependent phase shift is added to upper-channel directional parts of the wave  function to give
\begin{equation}
 \nonumber
 \Psi(\nu_1,\nu_2)\propto \exp\left[-\frac{(\nu_1+\nu_2)^2\tau_p^2}{2}\right]\exp\left[-\frac{(\nu_1-\nu_2)^2\tau_L^2}{2}\right]
\end{equation}
\begin{equation}
\times\left[\left(\begin{matrix}1\\0\end{matrix}\right)_1\left(\begin{matrix}0\\1\end{matrix}\right)_2 e^{i\nu_1\Delta t}+\left(\begin{matrix}0\\1\end{matrix}\right)_1\left(\begin{matrix}1\\0\end{matrix}\right)_2 e^{i\nu_2\Delta t}\right],
 \label{freq w.f.}
\end{equation}
where $\nu_{1,2}=\omega_{1,2}-\omega_0/2$ and $\tau_L=\sqrt{Lk_1^{\prime\prime}}/2$.

The temporal wave function is obtained from  $\Psi(\nu_1,\nu_2)$ with the help of the Fourier transformation
\begin{equation}
 \label{temporal}
 {\widetilde\Psi}(t_1,t_2)=\int d\nu_1d\nu_2\Psi(\nu_1,\nu_2)e^{i(\nu_1t_1+\nu_2t_2)}.
\end{equation}
The result of integrations is given by (with dropped unimportant phase factors)
\begin{equation}
 \nonumber
     {\widetilde\Psi}(t_1,t_2)\propto\exp\left[-\frac{(t_1+t_2+\Delta t)^2}{8\tau_p^2}\right]\times
\end{equation}
\begin{equation}
 \nonumber
\left\{\exp\left[-\frac{(t_1-t_2+\Delta t)^2}{8\tau_L^2}\right]
\left(\begin{matrix}1\\ 0\end{matrix}\right)_1\left(\begin{matrix}0\\ 1\end{matrix}\right)_2+
\right.
\end{equation}
\begin{equation}
 \label{integrated}
\left.\exp\left[-\frac{(t_1-t_2-\Delta t)^2}{8\tau_L^2}\right]
\left(\begin{matrix}0\\ 1\end{matrix}\right)_1\left(\begin{matrix}1\\ 0\end{matrix}\right)_2\right\}
\end{equation}
The time variables $t_1$ and $t_2$ can be interpreted as the arrival times of photons 1 and 2 to the beam splitter, occurring at zero delay time, $\Delta t=0$. Note that because of indistinguishability of photons their``numbers" 1 and 2 cannot be associated with numbers of channels $up$ and $down$. The numbers 1 and 2 indicate only numbers of photon variables, both temporal and directional, with the latter shown as subscripts at columns.

The first factor on the right-hand side of Eq. (\ref{integrated}) characterizes
dependence of the temporal wave function on the sum of two arrival times,
and it is identical for both terms in braces. For this reason it does not
affect relative distributions of photons between channels either before
or after propagation through BS, and below it's dropped.

As for the remaining part of the wave function ${\widetilde \Psi}$ (\ref{integrated}),
being transformed at BS, it takes the form
\begin{equation}
 \label{transf t1-t2)}
 {\widetilde \Psi}\propto A_+\Phi_-+A_-\Psi_- ,
\end{equation}
where $\Phi_--$ and $\Psi_-$ are the symmetric and
antisymmetric Bell states of Eqs. (\ref{Bell Phi-}) and (\ref{Bell Psi-}),
and $A_\pm$ are given by
\begin{equation}
 \label{A-pm}
  A_\pm =
e^{-\frac{(t_1-t_2+\Delta t)^2}{8\tau_L^2}}\pm
        e^{-\frac{(t_1-t_2-\Delta t)^2}{8\tau_L^2}}
\end{equation}
Note that the antisymmetric Bell state $\Psi_-$ on the right-hand side of Eq. (\ref{transf t1-t2)}) is multiplied by the function $A_-(t_1-t_2)$ which is also antisymmetric with respect to transposition $1\leftrightarrows 2$, which makes their product symmetric and not violating the general rule that the  wave function of two photons (bosons) is obliged to be symmetric with respect to transposition of all its variables.

\begin{figure}[h]
\centering
\includegraphics[width=5 cm]{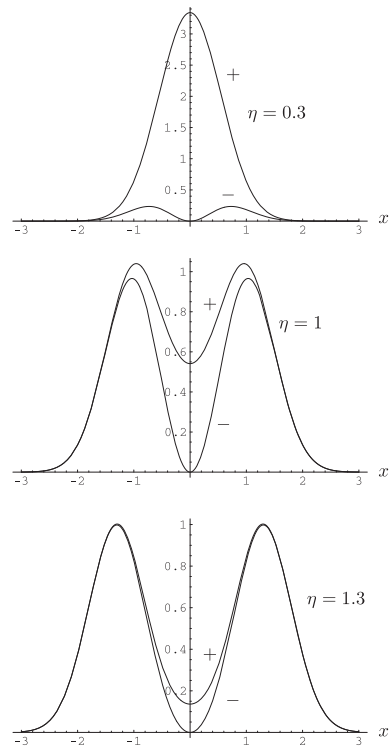}
\caption{{\protect\footnotesize {Probability densities (\ref{prob dens}) for biphotn pairs to be unsplit ($+$) and split ($-$) vs. the difference of arrival times ($t_1-t_2$).}}}\label{Fig3}
\end{figure}

The probability densities for photon pairs to be split or unsplit are proportional to the squared absolute values of the amplitudes
\begin{equation}
 \nonumber
 f_\pm(t_1-t_2)\equiv\frac{dw_\pm}{d(t_1-t_2)}\propto |A_\pm|^2=
\end{equation}
\begin{equation}
 \label{prob dens}
 \left|\exp\left[-\frac{(t_1-t_2+\Delta t)^2}{8\tau_L^2}\right]\pm
        \exp\left[-\frac{(t_1-t_2-\Delta t)^2}{8\tau_L^2}\right]\right|^2,
\end{equation}
where $+$ corresponds to unsplit and $-$ to split pairs.
Behavior of the functions $f_\pm(t_1-t_2)$ is determined mainly by the dimensionless control parameter $\eta=\Delta t/\sqrt{8}\tau_L$.  In Fig. \ref{Fig3} these functions are shown at $\eta=0.3,\,1\,{\rm and}\,1.3$ in dependence on the dimensionless variable $x=(t_1-t_2)/\sqrt{8}\tau_L$.

As it's seen, in the case of small values of $\eta$ ($\Delta t\ll\tau_L$) the probability density for photon pairs is much smaller than the probability for photon pairs to be unsplit, $f_-(t_1-t_2)\ll f_+(t_1-t_2)$, which means that that the HOM effect is close to the ideal one. In the opposite case, $\eta\gg 1$  ($\Delta t\gg\tau_L$) the curves $f_-(t_1-t_2)$ and  $f_+(t_1-t_2)$ are practically identical, which indicates missing interference and completely destroyed HOM effect. And, at last, in the intermediate case, at $\eta=1$ ($\Delta t\sim\tau_L$) the probability densities $f_+(t_1-t_2)$ and $f_-(t_1-t_2)$ differ from each other but do not coincide, which means that there are both split and unsplit photon pairs in comparable amounts.

Except for the case of a small delay time $\Delta t$, both curves $f_-(t_1-t_2)$ and  $f_+(t_1-t_2)$ have a double-peak structure. Positions of peaks correspond to such values of $t_1-t_2$, which compensate exactly the delay $\Delta t$ resulting from the pathway lengthening,

\noindent $(t_1-t_2)_{\rm peak}=\pm\Delta t$. As for the width of peaks $\delta(t_1-t_2)$, it does not depend of $\Delta t$ and equals approximately $\tau_L$. This discloses a fundamental feature of the process of pair production in SPDC. The two photons born simultaneously in a crystal not necessarily come to detectors exactly at the same time, because in a crystal they can move with not exactly coinciding velocities. In the case of the type-I phase matching and degenerate central frequencies which we consider here, the group velocities of two emitted photons are equal to each other. But owing to dispersion (the second-order derivative of the wave vector) there is some spreading of photon propagation velocities in a crystal. This effect shows itself in uncertainty of photon frequency difference $|\omega_1-\omega_2|\sim 1/\tau_L$ and related to it uncertainty of the arrival times $|t_1-t_2|\sim \tau_L $ at $\Delta t=0$ and $\delta(t_1-t_2)\sim\tau_L$ in the case $\Delta t\neq 0$. In measurements such difference of arrival times can remain not seen because of a relatively large value of the detector's temporal resolution exceeding $\delta(t_1-t_2)\sim\tau_L$. But detectors with better temporal resolution can disclose this effect of not obligatory exactly simultaneous arrivals to detectors of two photons in each single given SPDC pair, e.g., in the case $\Delta t=0$.

Note that in principle the split-pair probability density $f_-(t_1-t_2)$ can be measured experimentally via coincidence measurements between the $up$ and $down$ channels if the temporal resolution of detectors is significantly smaller than  $\tau_L$ and if detectors in $up$ and $down$ channels are tuned on at different times and only for a short interval of time in each of many repeated measurements. Alternatively, detectors in two channels can fix independently all events of coming photons, and information is saved in the computer, which then selects and counts only events separated by any given time difference $t_{up}-t_{down}$. This gives a single point at the curve $f_-(t_1-t_2)$, and so on.

\begin{figure}[h]
\centering
\includegraphics[width=6 cm]{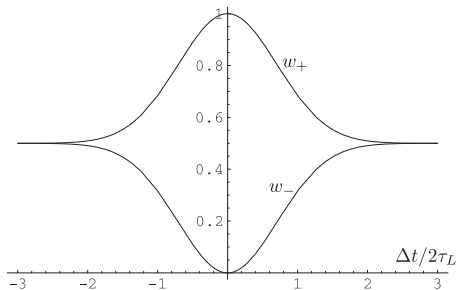}
\caption{{\protect\footnotesize {Total probabilities of getting unsplit ($w_+$) and split ($w_-$) photon pairs in dependence on the delay time $\Delta t$ related to the pathway lengthening in the upper channel.}}}\label{Fig4}
\end{figure}

If the temporal resolution of detectors is longer than both $\Delta t$ and $\tau_L$, the described detailed reconstruction of the probability densities is hardly possible, and one can watch only the total numbers of unsplit and split photon pairs determined by probabilities $w_\pm(\Delta t)$, which arise after integration of probability densities over both time variables $t_1$ and $t_2$. With the obvious normalization condition $w_++w_-=1$ taken into account, the result of integration can be presented in the form
\begin{equation}
 \label{total probab}
 w_\pm=
\frac{1}{2}\left\{1\pm\exp\left[-\left(\frac{\Delta t}{2\tau_L}\right)^2\right]\right\}.
\end{equation}
The curves $w_\pm(\Delta t)$ are shown in Fig. \ref{Fig4}
Both Eq. (\ref{total probab}) and the lower curve in Fig. \ref{Fig4} show that the ideal HOM effect occurs only at zero delay time $\Delta t=0$ and how it degrades with growing delay and with the scaling factor $\tau_L$. Asymptotically, at $\Delta t \gg \tau_L$ the HOM effect and interference disappear at all  and $w_+=w_-=\frac{1}{2}$.

\section{Conclusion}

We have considered two mechanisms capable to destroy the ideal interference HOM effect: difference of photon polarizations and the time delay of photons in one of the incidence channels. In the last case we used the biphoton wave function with two continuous frequency variables and related to it temporal wave function with two time variables. The time variables are interpreted as the arrival times of photons to the beam splitter or to the detectors. We believe that this approach is absolutely necessary for correct description of deviations from the ideal HOM effect arising owing to delays of photon arrivals to BS in one of two channels.

We have considered only a rather simple case of SPDC degenerate with respect to the central frequencies of emitted photons and only the case of relatively long pump pulses. Other cases and SPDC regimes can be even more interesting and they will be described elsewhere.


\section*{Acknowledgement}
The work is supported by the Russian Science Foundation, grant 14-12-01338$\Pi$.

\bibliography{Text}






\end{document}